\documentstyle[11pt,epsf]{article}

\newcommand{\be}{\begin{equation}}
\newcommand{\ee}{\end{equation}}
\newcommand{\bea}{\begin{eqnarray}}
\newcommand{\eea}{\end{eqnarray}}
\newcommand{\lap}{\mbox{}_{\textstyle \sim}^{\textstyle < }}

\newcommand{\pslash}{p\!\!/\,}

\begin{document}
\begin{titlepage}
\flushright{SCSC-TR-98-13}
\vspace{3cm}

\begin{center}
{\Large \bf Monte Carlo Quasi-Heatbath \\ by approximate inversion}

\vspace{1cm}
{Philippe de Forcrand}

{\small \it Swiss Center for Scientific Computing (SCSC) \\
\small \it  ETH-Z\"urich, CH-8092 Z\"urich \\
\small \it  Switzerland \\
{\bf forcrand@scsc.ethz.ch}
}
\end{center}
\vspace{2cm}

\abstract{When sampling the distribution 
$P(\vec{\phi}) \propto exp(-|A\vec{\phi}|^2)$,
a global heatbath normally proceeds by solving the linear system
$A\vec{\phi} = \vec{\eta}$,
where $\vec{\eta}$ is a normal Gaussian vector, exactly.
This paper shows how to preserve the distribution $P(\vec{\phi})$ while
solving the linear system with arbitrarily low accuracy.
Generalizations are presented. \\
}
\vspace{1cm}
PACS numbers: 02.70.L, 02.50.N, 52.65.P, 11.15.H, 12.38.G

\end{titlepage}


In Monte Carlo simulations, it is frequently the case that one wants to sample
a vector $\vec{\phi}$ from a distribution of the Gaussian type
$\propto \exp(-|A\vec{\phi}|^2)$. Typically, $\vec{\phi}$ has many components,
and $A$ is a large, sparse matrix. In lattice field theory, $\vec{\phi}$
is the value of the continuum field $\vec{\phi}$ at regular grid points,
and $A$ is the discretized version of some differential operator ${\cal A}$.
Illustrative examples used in this paper are 
${\cal A} = m + i \vec{p}$ (free field) and ${\cal A} = m + i \pslash$
(Dirac operator). The goal of the Monte Carlo simulation is to provide 
independent configurations of $\vec{\phi}$ at the least cost. 

The brute-force approach consists of drawing successive random vectors
$\vec{\eta}^{(k)}$ from the normal Gaussian distribution 
$\exp(-|\vec{\eta}|^2)$,
and of solving $A \vec{\phi}^{(k)} = \vec{\eta}^{(k)}$. The solution of 
this linear system can be efficiently obtained with an iterative linear solver
(Conjugate Gradient if $A$ is Hermitian, BiCGStab otherwise). This
approach can be called a global heatbath, because $\vec{\phi}^{(k+1)}$
has no memory of $\vec{\phi}^{(k)}$: the heatbath has touched all the 
components of $\vec{\phi}$. To avoid a bias, the solver must be iterated
to full convergence, which is often prohibitively expensive. One may try
to limit the accuracy while maintaining the bias below statistical errors,
but this requires a delicate compromise difficult to tune {\em a priori}.
A notable example of this global heatbath approach is the stochastic 
evaluation of inverse Dirac matrix elements,
where several hundred ``noise vectors'' $\vec{\eta}^{(k)}$ are inverted
to yield 
$(A^\dagger A)^{-1}_{ij} \approx \langle \phi_i \phi^\dagger_j \rangle_k$.
An abundant literature has been devoted to the optimization of this
procedure \cite{KFLiu,SESAM}.

For the free field or the Dirac operator mentioned above, the number of
iterations of the solver required to reach a given accuracy grows like the
correlation length $\xi \equiv 1/m$. Thus the work per new, independent
$\vec{\phi}$ is $c~\xi^z$ where $z$, the dynamical critical exponent,
is 1. However, the prefactor $c$ is large. For this reason, local updates,
where only one component of $\vec{\phi}$ is changed at a time, are often
preferred. They usually provide an independent $\vec{\phi}$ after an
amount of work $c'~\xi^2$, but with a much smaller prefactor $c'$ \cite{ft1}.
This paper presents an adaptation of the global heatbath which allows for
arbitrarily low accuracy in the solution of $A \vec{\phi} = \vec{\eta}$,
thus reducing the prefactor $c$, while maintaining the correct distribution.
This is obtained by the introduction of an accept/reject test of the
Metropolis type, making the procedure a ``quasi-heatbath.''
The method is described in the next section. Generalizations, including
a local version, are presented afterwards.

\section{Quasi-Heatbath}

Efficient Monte Carlo often relies on the subtle introduction of auxiliary
degrees of freedom. Consider here a vector $\vec{\chi}$ distributed 
according to $\frac{1}{Z_\chi} \exp(-|\vec{\chi} - A \vec{\phi}|^2)$.
Note that $Z_\chi$ is a constant ($\pi^N$ for an $N$-component complex
vector) independent of $\vec{\phi}$.
Therefore, the original distribution of $\vec{\phi}$, 
$\frac{1}{Z_\phi} \exp(-|A \vec{\phi}|^2)$,
is unchanged by the introduction of $\vec{\chi}$:
\be
\frac{1}{Z_\phi} \int {\cal D}\vec{\phi}~e^{-|A \vec{\phi}|^2} =
\frac{1}{Z_\phi Z_\chi} \int {\cal D}\vec{\phi} {\cal D}\vec{\chi}~
e^{-|A \vec{\phi}|^2 - |\vec{\chi} - A \vec{\phi}|^2}.
\label{eq1}
\ee
We can now alternate Monte Carlo steps on $\vec{\phi}$ and $\vec{\chi}$,
with the following prescription: 
\begin{enumerate}
\item Perform a global heatbath on $\vec{\chi}$: 
\be
\vec{\chi} \longleftarrow A \vec{\phi} + \vec{\eta},
\label{eq2}
\ee
where $\vec{\eta}$ is a normal Gaussian vector; \\
\item Reflect $A \vec{\phi}$ with respect to the minimum of the quadratic form
$(|A \vec{\phi}|^2 + |\vec{\chi} - A \vec{\phi}|^2)$: 
\begin{center}
$A \vec{\phi} \longleftarrow \vec{\chi} - A \vec{\phi}$,
\end{center}
i.e.,
\be
\vec{\phi} \longleftarrow A^{-1} \vec{\chi} - \vec{\phi}.
\label{star}
\ee
\end{enumerate}
Step 2 conserves the probability of $\vec{\phi}$ but is not ergodic.
Step 1 provides the ergodicity. Note that step 2 exchanges the two terms
$|A \vec{\phi}|^2$ and $|\vec{\chi} - A \vec{\phi}|^2$ in the quadratic form.
Since $\vec{\chi} - A \vec{\phi}$ in step 1 is set to a new random vector
$\vec{\eta}$, $A \vec{\phi}$ at the end of step 2 is equal to $\vec{\eta}$.
Therefore, a completely decorrelated $\vec{\phi}$ has been generated.
The vector $\vec{\chi}$ is not needed any longer and can be discarded.

This two-step algorithm can now be modified slightly. The vector 
$A^{-1} \vec{\chi}$ in Eq.(\ref{star}) need not be computed exactly.
Consider an approximate solution $\vec{\zeta}$ with 
$A \vec{\zeta} = \vec{\chi} - \vec{r}$, where $\vec{r} \neq \vec{0}$ 
is the residual. Step 2 should now be considered as a way to propose
a candidate $\vec{\phi'} = \vec{\zeta} - \vec{\phi}$ in a Metropolis
procedure. Since $\vec{\zeta}$ is completely independent of $\vec{\phi}$
or $\vec{\phi'}$, the probability of proposing $\vec{\phi'}$ given $\vec{\phi}$
is the same as that of proposing $\vec{\phi}$ given $\vec{\phi'}$.
Detailed balance will be satisfied with the additional step: 
\begin{enumerate}
\setcounter{enumi}{2}
\item Accept the candidate $\vec{\phi'} = \vec{\zeta} - \vec{\phi}$ with
probability
\be
P_{\rm acc}(\vec{\phi} \rightarrow \vec{\phi'}) = \mbox{min}(1,e^{-\Delta S}),
\label{Met}
\ee
where $\Delta S = |A \vec{\phi'}|^2 + |\vec{\chi} - A \vec{\phi'}|^2
- |A \vec{\phi}|^2 - |\vec{\chi} - A \vec{\phi}|^2$. 
\end{enumerate}
Simple algebra shows that 
\begin{center}
$\Delta S = 2 \mbox{Re}(r^\dagger \cdot (A \vec{\phi} - A \vec{\phi'}))$,
\end{center}
which is antisymmetric under the exchange 
$\vec{\phi} \leftrightarrow \vec{\phi'}$, as it should be.
If the linear system $A \vec{\zeta} = \vec{\chi}$ is solved exactly
($\vec{r} = \vec{0}$), then $\Delta S = 0$ and one recovers the original
global heatbath with acceptance 1.
Otherwise, the candidate $\vec{\phi'}$ may be rejected, in which case
$\vec{\phi}^{(k)}$ must be included once more in the Monte Carlo sequence:
$\vec{\phi}^{(k+1)} = \vec{\phi}^{(k)}$. As the residual is allowed to grow,
the average acceptance falls. But no bias is introduced: the distribution
of $\vec{\phi}$ remains $\frac{1}{Z_\phi} \exp(-|A \vec{\phi}|^2)$.

The optimal magnitude of $\vec{r}$ is thus the result of a compromise 
between accuracy and acceptance. The average acceptance of the prescription
(\ref{Met}) is $\mbox{erfc}(\sqrt{\langle (\Delta S)^2 \rangle / 8})$ \cite{ERFC}. 
Here $\langle (\Delta S)^2 \rangle$ can be evaluated as a function of the 
convergence criterion
$\epsilon$ of the linear solver. If the solver yields a residual $\vec{r}$
such that $\frac{||\vec{r}||}{||\vec{\chi}||} \leq \epsilon$, then
$\langle (\Delta S)^2 \rangle \leq 8 N \epsilon^2$,
where $A \vec{\phi}$, $A \vec{\phi'}$, and $\vec{r}$ have been considered
independent random Gaussian vectors with variance $N$, $N$, and $2 \epsilon^2 N$,
respectively, and $N$ is the number of their components.
Therefore, the acceptance is simply
\be
\langle \mbox{acceptance} \rangle \approx \mbox{erfc}(\epsilon \sqrt{N}).
\label{acc}
\ee
In other words, the acceptance is entirely determined by $\epsilon$ and $N$,
the number of degrees of freedom (the volume) of the system, and is
independent of the matrix $A$. To maintain a constant acceptance as the
volume grows, the convergence criterion for the solution of 
$A \vec{\zeta} = \vec{\chi}$ should vary like $1 / \sqrt{N}$.
An accuracy $\epsilon \sim 10^{-3} - 10^{-4}$ provides an acceptance of
80-90\% up to systems of $10^6$ degrees of freedom. There is no need for
higher accuracy.

The convergence of an iterative solver is typically exponential:
$\epsilon_n \sim e^{-n/\xi}$ after $n$ iterations. Therefore, the above prescription
reduces the work by a factor of 2 to 3 compared to the usual approach which
iterates the solver until ``full'' convergence (which typically means
$\epsilon~\lap~10^{-8} - 10^{-12}$). Illustrative results are shown in 
Fig. 1 for the case of the Wilson--Dirac operator. This figure shows the number of
iterations, the acceptance, and the work per independent $\vec{\phi}$ as a function of
$\epsilon$. The acceptance obeys $\mbox{erfc}(c~\epsilon \sqrt{N})$, where $c < 1$ 
(0.75 here)
reflects the fact that the residual is always smaller ($c$ times smaller on average)
than required by the stopping criterion. For this system of $N = 49\ 152$ variables 
($8^4$ lattice), the optimal stopping criterion is near $10^{-3}$.

\begin{figure}[ht]
\begin{center}
\mbox{\epsfxsize=10cm\epsfysize=10cm\epsffile[80 160 580 685]{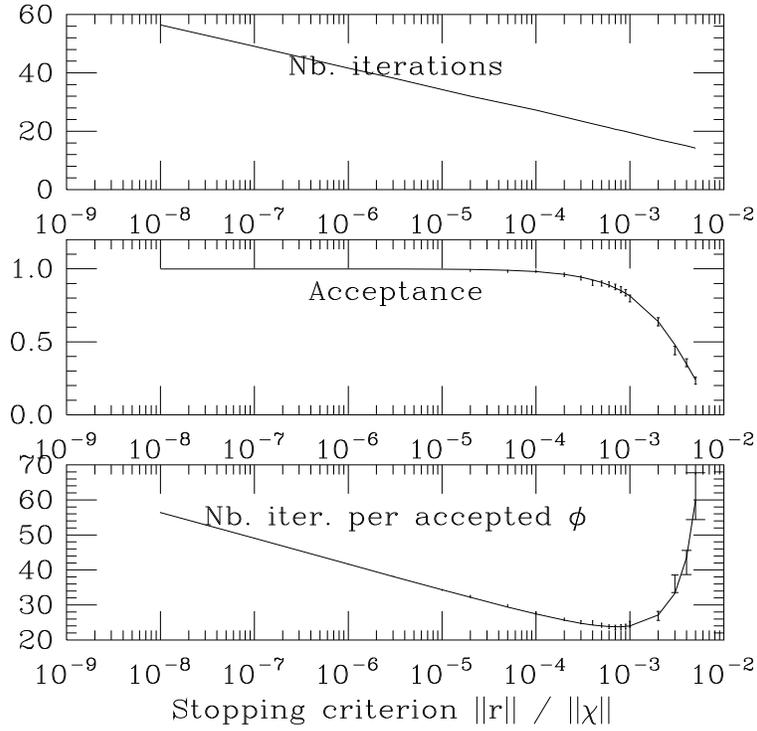}}
\end{center}
\caption{As the stopping criterion in the iterative solver is varied, the number of
solver iterations (top) and the acceptance of the quasi-heatbath (middle) change. 
The acceptance is well described by Eq.(\ref{acc}) (solid line). 
The work per new $\vec{\phi}$ (bottom) shows a clear minimum. 
The optimal stopping criterion depends on the system size only (49\ 152 here).}
\end{figure}

\section{Generalizations}

\subsection*{A. Over- and under-relaxation}

Consider a modification of Eq.(\ref{eq1}) with a parameter $\lambda$:
\be
\frac{1}{Z_\phi} \int {\cal D}\vec{\phi}~e^{-|A \vec{\phi}|^2} =
\frac{1}{Z_\phi Z_\chi} \int {\cal D}\vec{\phi} {\cal D}\vec{\chi}~
e^{-|A \vec{\phi}|^2 - |\vec{\chi} - \lambda A \vec{\phi}|^2}.
\label{eqlambda}
\ee

The same 3-step algorithm of Section 1 now reads:
\begin{enumerate}
\item Heatbath on $\vec{\chi}$: 
\be
\vec{\chi} \longleftarrow \lambda~A \vec{\phi} + \vec{\eta};
\label{eq3}
\ee
\item Reflection of $\vec{\phi}$ about the approximate minimum 
of the quadratic form:
\be
\vec{\phi'} = \frac{2 \lambda}{1 + \lambda^2} \vec{\zeta} - \vec{\phi},
\label{eq4}
\ee
where $A \vec{\zeta} = \vec{\chi} - \vec{r}$; \\
\item Accept $\vec{\phi'}$ with probability
$P_{\rm acc}(\vec{\phi} \rightarrow \vec{\phi'}) = \mbox{min}(1,e^{-\Delta S})$
where $\Delta S = |A \vec{\phi'}|^2 + |\vec{\chi} - \lambda A \vec{\phi'}|^2
- |A \vec{\phi}|^2 - |\vec{\chi} - \lambda A \vec{\phi}|^2$;
and, by simple algebra,
\be
\Delta S = 2 \lambda~\mbox{Re}(r^\dagger \cdot (A \vec{\phi} - A \vec{\phi'})).
\label{eq5}
\ee
\end{enumerate}

Thus, as $\lambda$ decreases from 1, $\langle (\Delta S)^2 \rangle$ also
decreases, which 
boosts the acceptance. On the other hand, Eq.(\ref{eq4}) indicates that
$\vec{\phi'}$ approaches $-\vec{\phi}$ as $\lambda \rightarrow 0$, so that 
$\vec{\phi'}$ and $\vec{\phi}$ become very (anti)correlated. 
The parameter $\lambda$ allows interpolation
between simple reflection ($\lambda = 0$) and no motion at all
($\lambda = +\infty$). In fact, substituting Eq.(\ref{eq3}) into Eq.(\ref{eq4})
gives 
\be
\vec{\phi'} = -\frac{1 - \lambda^2}{1 + \lambda^2} \vec{\phi} 
+ \frac{2 \lambda}{1 + \lambda^2} (A^{-1} \vec{\eta} - \vec{r}).
\label{eq6}
\ee
Taking $\vec{r} = \vec{0}$, one can identify this prescription with that of
Adler's stochastic over-relaxation (AOR) \cite{Adler}: \\
\be
\vec{\phi'} = (1 - \omega) \vec{\phi} 
+ \sqrt{\omega (2 - \omega)} A^{-1} \vec{\eta},
\ee
provided $\omega = \frac{2}{1 + \lambda^2}$. The quasi-heatbath can be viewed 
as a flexible, {\em global} generalization of Adler's AOR.

It is clear from Eq.(\ref{eq5}) that $\lambda < 1$ allows for a looser
convergence criterion $\epsilon \sim 1/\lambda$. However, the work to reach
convergence typically grows like $-\rm{log}~\epsilon$, whereas Eq.(\ref{eq6})
indicates that the number of Monte Carlo steps to decorrelate $\vec{\phi}$ will
grow like $\frac{1}{\lambda^2}$. Therefore, it seems inadvisable to depart
from $\lambda=1$.

Nonetheless, there are many situations where a completely independent 
$\vec{\phi}$ at each Monte Carlo step is a wasteful luxury. When the matrix
$A$ fluctuates and depends on other variables $U$, it will take some time
for the $U$ to equilibrate in the new background $\vec{\phi}^{(k+1)}$.
Equilibration will be achieved quickly over short distances, more slowly
over large ones. In that case it is useful to refresh the short-wavelength
modes of $\vec{\phi}$ at every MC step, but not the long-wavelength ones.
The situation is similar for the stochastic evaluation of inverse Dirac
matrix elements: one is interested in estimating $(A^\dagger A)^{-1}_{ij}$,
where the distance $|i-j|$ is short. Refreshing the long-wavelength modes
every time is wasteful.

\subsection*{B. Selective mode refresh}

The quasi-heatbath may be tailored for this purpose by modifying the basic
Eq.(\ref{eq1}) to 
\be
\frac{1}{Z_\phi} \int {\cal D}\vec{\phi}~e^{-|A \vec{\phi}|^2} =
\frac{1}{Z_\phi Z_\chi} \int {\cal D}\vec{\phi} {\cal D}\vec{\chi}~
e^{-|A \vec{\phi}|^2 - |\vec{\chi} - C \vec{\phi}|^2}.
\label{eqrefresh}
\ee

The matrix $C$ plays the role of the earlier $\lambda A$, 
except that now $\lambda$ depends on the eigenmode considered. 
The three basic steps of the algorithm become: 
\begin{enumerate}
\item Heatbath on $\vec{\chi}$: 
\be
\vec{\chi} \longleftarrow C \vec{\phi} + \vec{\eta};
\ee
\item Reflection of $\vec{\phi}$ about the approximate minimum 
of the quadratic form:
\be
\vec{\phi'} = \vec{\zeta} - \vec{\phi},
\ee
where 
\be
\frac{1}{2} (A^\dagger A + C^\dagger C) \vec{\zeta} = 
C^\dagger \vec{\chi} - \vec{r};
\label{eq7}
\ee
\item Accept $\vec{\phi'}$ with probability
$P_{\rm acc}(\vec{\phi} \rightarrow \vec{\phi'}) = \mbox{min}(1,e^{-\Delta S})$,
where 
\begin{center}
$\Delta S = 2 \mbox{Re}(r^\dagger \cdot (\vec{\phi} - \vec{\phi'}))$.
\end{center}
\end{enumerate}

For simplicity, consider the case where $C$ and $A$ commute. 
The candidate $\vec{\phi'}$ can be expressed as 
\begin{center}
$\vec{\phi'} = 
-(A^\dagger A + C^\dagger C)^{-1} (A^\dagger A - C^\dagger C) \vec{\phi} 
+ 2 (A^\dagger A + C^\dagger C)^{-1} (C^\dagger \vec{\eta} - \vec{r})$.
\end{center}
One wishes to obtain a heatbath ($\lambda \sim 1$ in Section A) 
on short-wavelength modes.
This implies a cancellation of eigenvalues in $(A^\dagger A - C^\dagger C)$ 
for short wavelengths. For long wavelengths a heatbath is not necessary,
and one could have $\lambda \sim 0$ or $+\infty$. 
One possible way to implement this would be
\begin{center}
$C = F^{-1} \Lambda F A$,
\end{center}
where $F$ is the Fourier transform and $\Lambda$ is a diagonal matrix
with entries $\lambda(\vec{k})$ growing from 0 to 1 with momentum $|\vec{k}|$. 
However, for operators ${\cal A}$ of the free-field or
Dirac type, a simpler and equivalent way consists of
modifying the mass parameter $m$
to $m_C > m$. This is equivalent to 
$\lambda(|\vec{k}|) = \sqrt{(m_C^2 + k^2)/(m^2 + k^2)}$.

The mass which enters into the linear system to solve (\ref{eq7})
is $m_{\rm eff} = \sqrt{(m^2 + m_C^2)/2}$.
As $m_C$ is increased, so is $m_{\rm eff}$. The work to approximately
solve Eq.(\ref{eq7}) decreases as $1 / m_{eff}$. Therefore, one achieves the
desired effect of refreshing short-wavelength modes at cheaper cost.
By drawing $m_C$ randomly from a suitable 
distribution at each MC step, the tailored refreshing of all
Fourier modes with the desired frequency can be achieved.

\subsection*{C. Local version}

The quasi-heatbath described so far is a global update procedure: all 
components of $\vec{\phi}$ are updated together. A local version readily
suggests itself:  restricting the auxiliary vector $\vec{\chi}$ to
have only 1 non-zero component, $\chi_i = \chi~\delta_{i,i_0}$ (or any subset
of components). 

Eq.(\ref{eq1}) then becomes 
\be
\frac{1}{Z_\phi} \int {\cal D}\vec{\phi}~e^{-|A \vec{\phi}|^2} =
\frac{1}{Z_\phi Z_\chi} \int {\cal D}\vec{\phi} {\cal D}\vec{\chi}~
e^{-|A \vec{\phi}|^2 - |\chi - (A \vec{\phi})_{i_0}|^2}.
\label{eqlocal}
\ee

The algorithm is unchanged: 
\begin{enumerate}
\item Heatbath on $\chi$:
$\chi \longleftarrow (A \vec{\phi})_{i_0} + \eta$; \\
\item Approximate reflection of $\vec{\phi}$:
$\vec{\phi'} = \vec{\zeta} - \vec{\phi}$, where
$A \vec{\zeta} = \vec{\chi} - \vec{r}$; \\
\item Accept $\vec{\phi'}$ with probability 
$P_{\rm acc}(\vec{\phi} \rightarrow \vec{\phi'}) = \mbox{min}(1,e^{-\Delta S})$, \\

where $\Delta S = \mbox{Re}(r^\dagger \cdot (A \vec{\phi} - A \vec{\phi'}))
+ \mbox{Re}(r_{i_0}^\star \cdot ((A \vec{\phi})_{i_0} - (A \vec{\phi'})_{i_0}))$.
\end{enumerate}

In this case, $\vec{\zeta}$ is the approximate Green's function of $A$
for a source at $i_0$. It will have a support of size ${\cal O}(\xi)$,
so that the local change in $\chi_{i_0}$ will induce a change in $\phi$
over a whole domain. By varying $i_0$ from 1 to $N$, one sweeps the whole
system and generates a new vector $\vec{\phi}^{(k+1)}$. If the acceptance
is maintained close to 1, $\vec{\phi}^{(k+1)}$ will essentially be
uncorrelated with $\vec{\phi}^{(k)}$. However, the work per local update 
is proportional to $\xi^d$ in $d$ dimensions, so that this approach 
becomes very inefficient
when the correlation length $\xi$ is large. Nevertheless, it may be
advantageous for moderate $\xi$. The reason is that the approximate
solution $\vec{\zeta} \approx A^{-1} \chi \delta(i_0)$ need not be
obtained by a Krylov method, which applies successive powers of $A$
to the initial residual. Instead, one may search for the best solution
$\vec{\zeta}$ among all vectors of localized support, for instance,
$i_0$ and its nearest neighbours.

\subsection*{D. Adler's stochastic over-relaxation}

Finally, the local variable $\chi~\delta(i,i_0)$ may interact with 
$\vec{\phi}$ in the simplest way, with a contact interaction.  This modifies
Eq.(\ref{eq1}) to
\be
\frac{1}{Z_\phi} \int {\cal D}\vec{\phi}~e^{-|A \vec{\phi}|^2} =
\frac{1}{Z_\phi Z_\chi} \int {\cal D}\vec{\phi} d\chi~
e^{-|A \vec{\phi}|^2 - |\chi - \lambda \phi(i_0)|^2}.
\ee
If one chooses to update only $\phi(i_0)$ and leave the other components
of $\vec{\phi}$ unchanged, then there is no need to invert the matrix $A$.
The algorithm simplifies to: 
\begin{enumerate}
\item Heatbath on $\chi$:
$\chi \longleftarrow \lambda \phi(i_0) + \eta$; \\
\item Reflection of $\phi(i_0)$ with respect to the minimum of the quadratic
form 
\begin{center}
$(m^2 + \lambda^2) |\phi(i_0)|^2
+ (\phi(i_0)^\dagger \cdot (\psi - \lambda \chi) + \mbox{H.c.})
+ \mbox{constant}$,
\end{center}
where $m^2 \equiv (A^\dagger A)_{i_0 i_0}$ and
$\psi \equiv (A^\dagger A)_{i_0 j} \phi(j)$.
\end{enumerate}
This reflection is exact, and so the acceptance test disappears. The new reflected
value is 
\begin{eqnarray*}
\phi'(i_0) & = & 2~\frac{\lambda \chi - \psi}{1 + \lambda^2} - \phi(i_0) \\
           & = & - \frac{1 - \lambda^2}{1 + \lambda^2} \phi(i_0)
- \frac{2}{1 + \lambda^2} \psi + \frac{2 \lambda}{1 + \lambda^2} \eta .
\end{eqnarray*}
This prescription is identical to Adler's stochastic over-relaxation
\cite{Adler} with the change of notation 
$\omega \leftrightarrow 2 / (1 + \lambda^2)$.

\section{Conclusion}

The quasi-heatbath Eqs.(\ref{eq2})--(\ref{Met}) is a simple and efficient
method to globally change a vector $\vec{\phi}$ distributed according to
$\frac{1}{Z_\phi} e^{-|A \vec{\phi}|^2}$. Like the global heatbath 
consisting of solving $A \vec{\phi} = \vec{\eta}$, where $\vec{\eta}$ is
a Gaussian vector, exactly at each Monte Carlo step, the quasi-heatbath
also has dynamical critical exponent 1. The prefactor is reduced by
a factor of 2 to 3 because the linear system $A \vec{\phi} = \vec{\eta}$ can
now be solved approximately. Whatever the level of accuracy, an acceptance
test maintains the exact distribution $e^{-|A \vec{\phi}|^2}$. The most
efficient choice for the accuracy level is ${\cal O}(1/\sqrt{N})$, where
$N$ is the volume of the system.

Several generalizations of the quasi-heatbath have been proposed. A simple
modification makes it possible to refresh each of the Fourier components of $A \vec{\phi}$
at a prescribed rate. A local version may be advantageous when the 
correlation length is moderate. In a limiting case, this version
becomes identical to Adler's stochastic over-relaxation.

\vspace{1cm}
I thank Massimo D'Elia for interesting discussions and valuable comments.


\begin{thebibliography}{99}
\bibitem{ft1} Adler's stochastic over-relaxation \cite{Adler} in principle 
allows $z=1$ to be reached. 
In practice, this requires careful tuning of the over-relaxation
coefficient depending on the spectrum of $A$.  This is usually difficult
to achieve, especially when $A$ itself is fluctuating.
\end{thebibliography}
\end{document}